\documentclass{emulateapj}

\usepackage{float}
\usepackage{amsmath}
\usepackage{epsfig,floatflt}
\usepackage{subfigure}

\begin{document}

\title{The two- and three-point correlation functions of the polarized five-year
{\it WMAP} sky maps}

  \author{E. Gjerl\o w\altaffilmark{1,6}, 
  H.\ K.\ Eriksen\altaffilmark{2,6,7},
  A.\ J.\ Banday\altaffilmark{3,8,9}, K.\ M.\ G\'orski\altaffilmark{4,10,11}
   and P.\ B.\ Lilje\altaffilmark{5,6,7}}

\altaffiltext{1}{Electronic address: eirik.gjerlow@astro.uio.no}
\altaffiltext{2}{Electronic address: h.k.k.eriksen@astro.uio.no}
\altaffiltext{3}{Electronic address: Anthony.Banday@cesr.fr}
\altaffiltext{4}{Electronic address: Krzysztof.M.Gorski@jpl.nasa.gov}
\altaffiltext{5}{Electronic address: per.lilje@astro.uio.no}
\altaffiltext{6}{Institute of Theoretical Astrophysics, University of Oslo,
P.O.\ Box 1029 Blindern, N-0315 Oslo, Norway}
\altaffiltext{7}{Centre of Mathematics for Applications,
University of Oslo, P.O.\ Box 1053 Blindern, N-0316 Oslo, Norway}
\altaffiltext{8}{Centre d'Etude Spatiale des Rayonnements
9, av du Colonel Roche, BP 44346 31028 Toulouse Cedex 4, France} 
\altaffiltext{9}{Max-Planck Institute f{\"u}r Astrophysik, Karl-Schwarzschild Str. 1, D-85748, Garching, Germany}
\altaffiltext{10}{JPL, M/S 169/327, 4800 Oak Grove Drive,
  Pasadena CA 91109, USA} 
\altaffiltext{11}{Warsaw University Observatory, Aleje
Ujazdowskie 4, 00-478 Warszawa, Poland}

%\date{Received - / Accepted -}

\begin{abstract}
We present the two- and three-point real space correlation functions
of the five-year {\it WMAP} sky maps and compare the observed functions to
simulated $\Lambda$CDM concordance model ensembles. In agreement with
previously published results, we find that the temperature correlation
functions are consistent with expectations. However, the pure
polarization correlation functions are acceptable only for the 33GHz
band map; the 41, 61, and 94 GHz band correlation functions all exhibit
significant large-scale excess structures. Further, these excess
structures very closely match the correlation functions of the two
(synchrotron and dust) foreground templates used to correct the {\it WMAP}
data for galactic contamination, with a cross-correlation
statistically significant at the $2\sigma-3\sigma$ confidence level. The
correlation is slightly stronger with respect to the thermal dust
template than with the synchrotron template. 
\end{abstract}


\maketitle

\section{Introduction}

%During the last two decades, cosmology has evolved from a data-starved
%to a data-driven branch of physics. The primary force behind this
%transition has been the development of ever more sensitive measuring
%devices, and in particular, more sensitive microwave detectors. As a
%result, experiments have been able to observe the very faint
%electromagnetic radiation created during the early universe known as
%the cosmic microwave background (CMB) with ever increasing accuracy.

Observations of the cosmic microwave background (CMB) have been among
the most important ingredients in the revolution of cosmology that has
taken place in the last two decades, when cosmology changed from a
data starved to a data rich science. The so far most influential
observations have been made with the {\it Wilkinson Microwave
Anisotropy Probe} ({\it WMAP}; Bennett et al. 2003a; Hinshaw et
al. 2007, 2009) satellite experiment, which has measured the microwave
sky at five different frequencies (23, 33, 41, 61, and 94 GHz) in both
temperature and polarization. Its main results are five sky maps with
resolutions between $13'$ and $55'$, each comprising more than three
million pixels in each of the three Stokes parameters I, Q, and U.

%Based on this and other data sets, theorists have today been able to
%establish a concordance $\Lambda$CDM model which can fit almost all
%currently available observations with only six free parameters. In
%addition to those six parameters, this model also requires, or
%predicts, the universe to be statistically isotropic and homogeneous,
%and filled with random Gaussian fluctuations. With a few notable
%exceptions, this picture is also well supported by available
%data. Clearly, cosmology as a whole has been a highly successful
%enterprise during the last few decades.

%This success would not have been possible if the improved detector
%technology had not been accompanied by improved computer technologies
%and algorithms. Only during the last ten years, the number of pixels
%available in a CMB sky map has increased from a few thousands to
%several millions. And in only a few years from now, this will increase
%further to tens or hundreds of million of pixels, when Planck releases
%its data products.

The detector technology leading to 
instruments like {\it WMAP} has thus been very important in observing the CMB.
Almost equally important has been the progress in computer
technology and algorithms. With the enormously increased number of
pixels from new high-resolution instruments, scientists have had to
develop clever algorithms for every step of the required data analysis
pipeline: map making (e.g., Ashdown et al. 2007; Hinshaw et al. 2003b,
and references therein), component separation (e.g., Bennett et
al. 2003b; Leach et al. 2008; Eriksen et al. 2008, and references
therein), power spectrum and cosmological parameter estimation (e.g.,
G{\'o}rski 1994; Lewis \& Bridle 2002; Hivon et al. 2002; Hinshaw et
al. 2003a; Verde et al. 2003; Eriksen et al. 2004d, and references
therein), and analysis of higher-order statistics (e.g., Hinshaw et
al. 1994; Kogut et al. 1995; Komatsu et al. 2003; Eriksen et
al. 2004c, 2005 and references therein).

In this paper, we revisit a well-known example of the latter category,
namely, real-space $N$-point correlation functions. These functions
arise naturally in studies of non-Gaussianity, since it can be shown
that any odd-ordered $N$-point function has an identically vanishing
expectation value, while all even-ordered $N$-point functions have an
expectation value given by products of the corresponding two-point
functions. Violation of either of these relations would indicate the
presence of a non-Gaussian component in the field under consideration.

Earlier $N$-point correlation function analyses of CMB data include
studies of the {\it COBE}-DMR data \citep{hinshaw:1995, kogut:1996,
  eriksen:2002}, the {\it WMAP} temperature data \citep{eriksen:2005} and
one single two-point analysis of the first-year {\it WMAP} polarization data
\citep{kogut:2003}. This paper is the first to consider the much more
mature five-year {\it WMAP} polarization data, and the first to compute the
three-point correlation function from any CMB polarization data. To do
this, we adopt and extend the $N$-point algorithms developed by
\citet{eriksen:2004b}.
 
\section{Methods and definitions}

\subsection{$N$-point correlation functions}
\label{sec:correlationfunctions}

An $N$-point correlation function $C_N$ of a stochastic field
$X(\hat{n})$ is defined as
\begin{equation}
C_N(\theta_1, \theta_2, \ldots, \theta_k) =
\biggl<\textrm{X}(\hat{n}_1)\textrm{X}(\hat{n}_2) \cdots  
\textrm{X}(\hat{n}_N)\biggr>,
\label{eq:Npointfunc}
\end{equation}
where $\{\hat{n}_{1}, \ldots, \hat{n}_{N}\}$ spans an $N$-polygon
defined by $k$ parameters, $\{\theta_1, \ldots, \theta_k\}$. In the
most general case in which no assumptions are made concerning the
statistical properties of the field $X$, one needs $k=2N$ parameters
to uniquely describe such a polygon\footnote{In this paper, we
  restrict our interest to fields defined on the two-dimensional
  sphere.}, namely, the individual positions of each vertex of the
polygon. However, in many applications one assumes that $X$ is
isotropic and homogeneous, and one can therefore average over
position. In such cases, the number of parameters are reduced by three on
a two-dimensional surface, corresponding to translation and
rotation. 

\begin{figure}[t]
\centering
\epsfig{figure=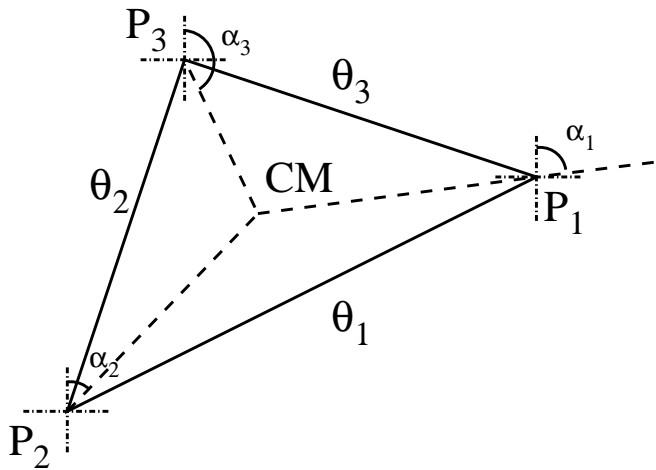,width=\linewidth}
\caption{Center of mass and rotation angles of a triangle. Each point $i$
is rotated by an angle $\alpha_i$ from the global coordinate system into the
local system defined by the triangle's center of mass.}
\label{fig:coordinates}
\end{figure} 

In this paper, we will consider only two- and three-point correlation
functions defined on the two-dimensional sphere, and we therefore need,
respectively, one and three parameters to describe our polygons (ie., line
and triangle). In the two-point case, the only natural parameter choice
is the angular distance, $\theta=\arccos(\hat{n}_1\cdot\hat{n}_2)$
between the two points, while in the three-point case there is some
freedom to choose. For simplicity, we parameterize the triangle by the
lengths of the three edges, $(\theta_1, \theta_2, \theta_3)$, where the edges
are ordered such that $\theta_1$ corresponds to the longest of the three
edges, and the remaining are listed according to clockwise traversal
of the triangle.

\subsection{Polarized correlation functions}

Our goal in this paper is to measure the $N$-point correlation
functions of the polarized {\it WMAP} maps. To do so, we have to generalize
the methods described by \citet{eriksen:2004b}, in order to account
for the fact that the CMB polarization field is a spin-2
field. Explicitly, the full CMB fluctuation field may be described in
terms of the three Stokes parameters, I($\hat{n}$), Q($\hat{n}$), and
U($\hat{n}$). Here I is the usual (scalar) temperature fluctuation
field, and Q and U are two parameters describing the linear
polarization properties of the radiation in direction $\hat{n}$.

According to standard CMB conventions, Q and U are defined with
respect to the local meridian of the spherical coordinate system of
choice. However, Q and U form a spin-2 field, and this means that if
one performs a rotational coordinate system transformation, the
Stokes parameters after transformation becomes
\begin{equation}
  \label{eq:qurot}
  \begin{pmatrix} \textrm{I'} \\\textrm{Q}' \\ \textrm{U}' \end{pmatrix} = 
\begin{pmatrix} 1 & 0 & 0 \\ 0 & \cos 2\alpha & \sin 2\alpha \\
               0& -\sin 2\alpha & \cos 2\alpha \end{pmatrix}
\begin{pmatrix} \textrm{I} \\ \textrm{Q} \\ \textrm{U} \end{pmatrix},
\end{equation}
where $\alpha$ is the local rotation angle between the two coordinate
systems. \citet{zaldarriaga:1997} and \citet{kamionkowski:1997} give
complete descriptions of the statistics of these fields.

\begin{figure*}[t]
\epsfig{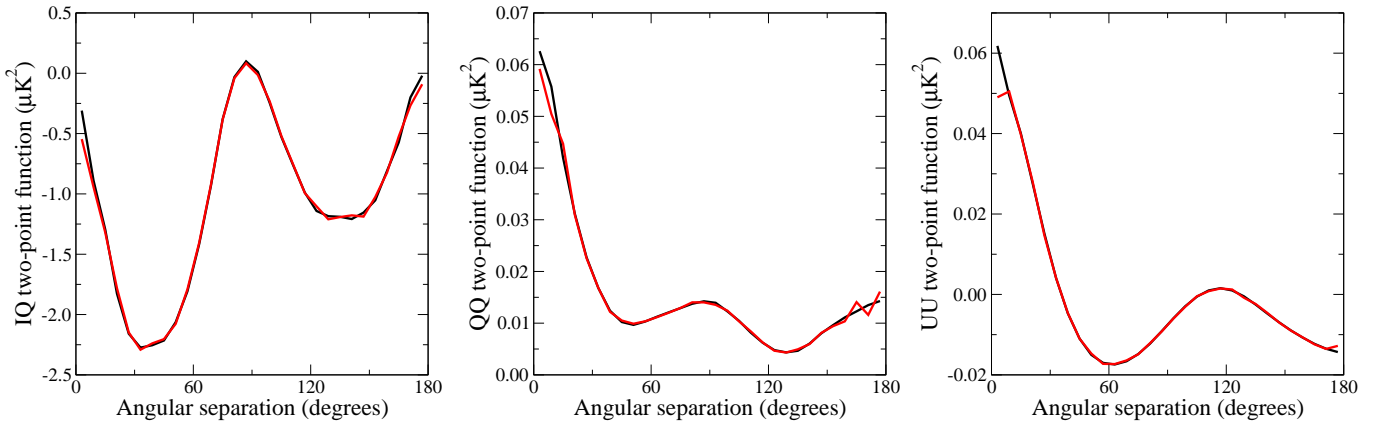}%, totalheight=\textheight}
\caption{Comparison between the analytically calculated two-point
  function (black) and the two-point function computed from the map
  with the corresponding $C_l$'s (red). The shown functions are, from
  left to right, the IQ, QQ, and UU correlation functions.}
\label{fig:theoryvsexp}
\end{figure*}

This property has some important consequences when computing $N$-point
correlation functions: since we assume that the CMB field is
isotropic and homogeneous, the value we obtain for a given element of
the correlation function should not depend on the coordinate
system. One cannot therefore blindly adopt Q and U as the quantities
to correlate, since these are in fact coordinate dependent. 

For the two-point function, this problem is conventionally solved by
defining a local coordinate system in which the local meridian
passes through the two points of interest
\citep{kamionkowski:1997}. The Stokes parameters in this new
``radial'' system are denoted by Q$^r$ and U$^r$, and measure the
polarization either orthogonally or perpendicularly (Q$^r$) to the
connecting line, or 45$^{\circ}$ rotated with respect to it
(U$^{r}$). The two-point function defined in terms of Q$^r$ and U$^r$
then becomes coordinate system independent.

We now generalize this idea to higher-order $N$-point correlation
functions. As with the definition of the $N$-point polygon parameters,
we also here have some freedom when choosing the reference points for
the coordinate independent quantities. For instance, two natural
choices for Q$^r_i$ and U$^r_i$ are to define them either with respect
to the edges or with respect to the center of mass of the polygon,
\begin{equation}
\hat{n}_{\textrm{CM}} = \frac{\sum_i \hat{n}_i}{|\sum_i \hat{n}_i|}.
\end{equation}
In this paper we choose the latter definition.
The resulting geometry is illustrated in Figure \ref{fig:coordinates}.

Given these new rotationally invariant quantities, $\textrm{X} \in
\{\textrm{I}, \textrm{Q}^r, \textrm{U}^r\}$, defined with respect to
the local center of mass, the polarized correlation functions are
simply defined as
\begin{equation}
C_N(\theta_1, \theta_2, \ldots, \theta_k) =
\biggl<\textrm{X}_1(\hat{n}_1)\textrm{X}_2(\hat{n}_2) \cdots  
\textrm{X}_N(\hat{n}_N)\biggr>,
\label{eq:pol_Npointfunc}
\end{equation}
Note that in the case of the two-point function, there are six
independent polarized correlation functions (II, IQ$^r$, IU$^r$,
Q$^r$Q$^r$, Q$^r$U$^r$ and U$^r$U$^r$), while in the three-point
function case, there are 27 independent components (III, IIQ$^r$,
$\ldots$, U$^r$U$^r$Q$^r$, U$^r$U$^r$U$^r$).

Algorithmically, we compute these functions from a given pixelized sky
map, $m(\hat{n})$, simply by averaging the corresponding products over
all available pixel multiplets that satisfy the geometrical
constraints of the polygon under consideration,
\begin{equation}
\hat{C}_N(\theta_1, \theta_2, \ldots, \theta_k) = \frac{1}{N_p}
\sum_{i}^{N_p} m^{X_1}_{p_1(i)} m^{X_2}_{p_2(i)} \cdots m^{X_N}_{p_N(i)}.
\end{equation} 
Here $i = 1, \ldots, N_p$ runs over the number of available pixel
multiplets, and $p_j(i)$ is the $j$'th pixel in the $i$'th pixel
multiplet. For full details on how to find the relevant pixel
multiplets corresponding to a given polygon efficiently, see
\citet{eriksen:2004b}.

The correlation functions are binned with a bin size tuned to the
pixel size of the map. Explicitly, since we use a HEALPix resolution
of $N_{\textrm{side}}=16$, with $\sim3^{\circ}$ pixels, we adopt a bin
size of $6^{\circ}$, for a total of 30 bins between $0^{\circ}$ and
$180^{\circ}$.

If all factors in an $N$-point multiplet are taken from the same map,
the resulting function is named an auto-correlation; otherwise, it is
called a cross-correlation. The main advantage of cross-correlations
is that the noise is typically uncorrelated between maps, while the
signal (ideally) is strongly correlated. Therefore, the noise
contribution to a cross-correlation averages to zero. For this reason,
one typically uses auto-correlations as a probe of systematic errors
(e.g., to check whether the assumed noise model is correct), but
cross-correlations for a final cosmological analysis. In this paper,
we consider both types.

In Figure \ref{fig:theoryvsexp}, we show a comparison of three two-point
correlation functions computed from the same signal-only simulation,
using two different methods: the red curves show the correlation
functions computed from the power spectrum of the realization,
employing the analytic expression given by, e.g.,
\citet{smith:2006}. The black curves show the correlation functions
computed directly from the pixelized map using the algorithms
described above. Slight differences are expected due to different
treatment of pixel windows, but clearly the agreement between the two
are excellent, giving us confidence that our machinery works as expected.

\subsection{Comparison with simulations}

In this paper we perform a standard frequentist analysis, in the sense
that we compare the results obtained from the real data with an
ensemble of simulations based on some model. Specifically, our
null-hypothesis here is that the universe is isotropic and
homogeneous, and filled with Gaussian fluctuations drawn from a
$\Lambda$CDM power spectrum. Our ensemble consists of
$N_{\textrm{sim}} = 100\,000$ simulations.

\begin{figure*}[t]
\centering
\epsfig{figure=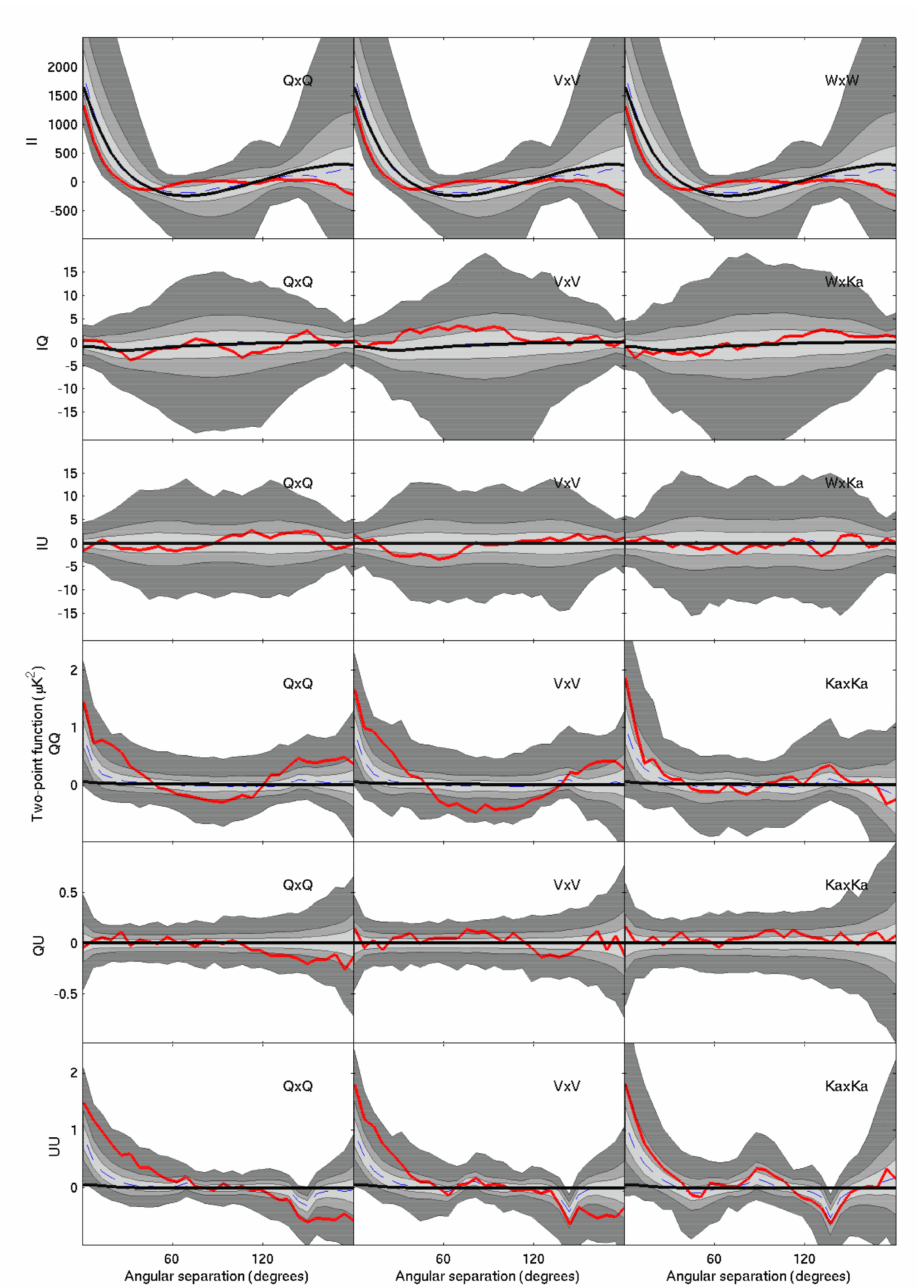, width=0.9\linewidth}%, totalheight=\textheight}
\caption{Two-point functions computed with the KQ85 mask,
as functions of angular
separation on the sky. Only the $Ka$, $Q$ and $V$ auto-correlations,
as well as the cross-correlations between the $W$ intensity map and the $Ka$
polarization map,
are shown. The red line shows the two-point function
computed from the {\it WMAP} five-year co-added maps. 
The light, medium and dark gray shaded areas correspond to 1$\sigma$, 
2$\sigma$, and 3$\sigma$, respectively. The dashed blue line is the maximal
point of the correlation histograms, while the black line is the theoretically
predicted (no noise) two-point function computed from the best-fit
five-year {\it WMAP}
$\Lambda$CDM power spectrum. Each row corresponds to a specific
combination of modes, while the frequency bands involved are marked on each 
plot.}
\label{fig:twopt_auto}
\end{figure*}

For each realization in the ensemble, we compute the $N$-point
correlation functions in precisely the same manner as for the real
data. Finally, we compare the correlation function derived from the
data with the simulated functions through a standard $\chi^2$
statistic,
\begin{equation}
  \label{eq:chisquared}
  \chi^2 = \sum_{b=1}^{N_{\textrm{bin}}}\frac{(C(b)-\mu(b))^2}{\sigma_{b}^2},
\end{equation}
where the sum runs over all $N$-point configurations/bins, and the mean
$\mu(b) = \bigl<C(b)\bigr>$ and variance $\sigma^2(b)=
\bigl<(C(b)-\mu(b))^2\bigr>$ are derived from the simulations. 

However, before computing the $\chi^2$ as described above, we
Gaussianize the distribution of each bin as follows
\citep{eriksen:2005}:
\begin{equation}
\frac{R}{N_{\textrm{sim}}+1}
= \frac{1}{\sqrt{2\pi}} \int_{-\infty}^{s}
e^{-\frac{1}{2} t^2} dt,
\label{eq:gaussianize2}
\end{equation}
where $R$ is the rank of the value under consideration (i.e., the
number of simulations with a \emph{lower} value than the currently
considered), and $s$ is the corresponding Gaussianized rank.  The
reason for performing this transformation is that the probability
distribution for a given bin is highly non-Gaussian, and for
even-ordered correlation functions strongly skewed toward high
values. A direct $\chi^2$ evaluation will therefore tend to give too
much weight to fluctuations that have  high values compared to fluctuations
that have low values. By first explicitly Gaussianizing, a symmetric
response is guaranteed.

\begin{deluxetable*}{cccccccccc}[t]
  \tablewidth{\linewidth}
  \tablecaption{Statistical significances (two-point functions) \label{tab:twopt}}
  \tablecomments{Statistical significances for the two-point functions as
  computed from the {\it WMAP} data. For the $W$ differencing assemblies, the
  first-year maps were used. For the band data, the five-year co-added maps were used.
  The Monte Carlo uncertainty is 1\%. 
  The correlations using both the KQ85 and the KQ75 masks are shown. As the
  pure polarization correlations are insensitive to which I mask is being
  used, these are only displayed under the KQ85 header.
  An entry with '\dots' signifies that
  some other entry in the table contains the same information as
  that entry, while an entry with '---' indicates that the information in 
  that entry was not computed.
  The listed ratios indicate the fraction of simulations with
  a higher $\chi^2$ than the observed data. The entries marked with an '*' 
  signifies that the fraction is lower than the Monte Carlo uncertainty, but
  still larger than zero; 
  consequently, the number of simulations (out of 100,000) 
  with a higher $\chi^2$ than the
  observed data are shown instead for these entries. The entries in parenthesis
  are the significances of the noise-only hypothesis.}
  \tablecolumns{10}
  \tablehead{& \multicolumn{9}{c}{Correlation} \\Frequency & II & IQ & IU & QI & QQ & QU & UI & UQ & UU}
  \startdata
  \cutinhead{KQ85}
   $Ka$x$Ka$ & --- & --- & --- & ---& 0.40 (0.32) & 0.45 (0.38) & ---& \dots & 0.32 (0.25) \\
   $Q$x$Q$ & 0.08 & 0.85 & 0.65 & \dots & 0.01 (348$^*$) & 0.01 (0.01) & \dots & \dots & 254$^*$ (128$^*$) \\
   $Q$x$V$ & 0.07 & 0.26 & 0.61 & 0.85 & 25$^*$ (18$^*$) & 0.06 (0.04) & 0.69 & 0.01 (338$^*$) & 0.04 (0.02) \\
   $Q$x$Ka$ & --- & 0.81 & 0.97 & --- & 0.01 (307$^*$) & 0.02 (0.01) & --- & 0.03 (0.02) & 154$^*$ (48$^*$) \\
   $Q$x$W$ & 0.08 & 0.84 & 0.05 & 0.86 & 3$^*$ (5$^*$) & 0$^*$ (0$^*$) & 0.69 &0$^*$ (0$^*$) & 253$^*$ (222$^*$) \\
   $V$x$V$ & 0.07 & 0.27 & 0.60 & \dots & 73$^*$ (49$^*$) & 0.02 (0.01) & \dots & \dots & 0.01 (393$^*$) \\
   $V$x$Ka$ & --- & 0.82 & 0.97 & --- & 0.01 (357$^*$) & 21$^*$ (15$^*$) & --- & 18$^*$ (10$^*$) & 0.02 (0.02) \\
   $V$x$W$ & 0.07 & 0.84 & 0.05 & 0.27 & 1$^*$ (2$^*$) & 42$^*$ (14$^*$) & 0.57 & 37$^*$ (8$^*$) & 106$^*$ (88$^*$) \\
   $W$x$Ka$ & --- & 0.78 & 0.98 & --- & --- & --- & --- & --- & --- \\
   $W$x$W$ & 0.07 & 0.83 & 0.05 & \dots & 3$^*$ (0$^*$) & 71$^*$ (34$^*$) & \dots & \dots & 2$^*$ (0$^*$) \\
   $W1$x$W1$ & 0.09 & 0.70 & 0.11 & \dots &116$^*$ (123$^*$) & 0.01 (0.01) & \dots & \dots & 5$^*$ (6$^*$) \\
   $W2$x$W2$ & 0.07 & 0.32 & 0.22 & \dots & 95$^*$ (83$^*$) & 0.03 (0.03) & \dots & \dots & 478$^*$ (476$^*$) \\
   $W3$x$W3$ & 0.07 & 0.91 & 0.82 & \dots & 0.78 (0.78) & 0.94 (0.94) & \dots & \dots & 0.60 (0.60) \\
   $W4$x$W4$ & 0.10 & 0.53 & 0.03 & \dots & 0$^*$ (0$^*$) & 0.08 (0.08) & \dots & \dots & 1$^*$ (3$^*$) \\
  \cutinhead{KQ75}
   $Q$x$Q$ & 0.07 & 0.96 & 0.61 & \dots & \dots & \dots & \dots & \dots & \dots \\
   $Q$x$V$ & 0.06 & 0.30 & 0.67 & 0.97 & \dots & \dots & 0.62 & \dots & \dots \\
   $Q$x$Ka$ & --- & 0.87 & 0.95 &  --- & \dots & \dots & --- & \dots & \dots \\
   $Q$x$W$ & 0.06 & 0.91 & 0.15 & 0.97 & \dots & \dots & 0.64 & \dots & \dots \\
   $V$x$V$ & 0.06 & 0.31 & 0.65 & \dots &\dots & \dots & \dots & \dots & \dots \\
   $V$x$Ka$ & --- & 0.88 & 0.96 & --- & \dots & \dots & --- & \dots & \dots \\
   $V$x$W$ & 0.06 & 0.92 & 0.16 & 0.32 & \dots & \dots & 0.62 & \dots & \dots \\
   $W$x$Ka$ & --- & 0.85 & 0.96 & ---& ---& ---& ---& ---& ---\\
   $W$x$W$ & 0.06 & 0.92 & 0.15 & \dots & \dots & \dots & \dots & \dots & \dots \\ 
   $W1$x$W1$ & 0.08 & 0.68 & 0.16 & \dots & \dots & \dots & \dots & \dots & \dots \\
   $W2$x$W2$ & 0.06 & 0.33 & 0.35 & \dots & \dots & \dots & \dots & \dots & \dots \\
   $W3$x$W3$ & 0.06 & 0.93 & 0.94 & \dots & \dots & \dots & \dots & \dots & \dots \\
   $W4$x$W4$ & 0.10 & 0.57 & 0.10 & \dots & \dots & \dots & \dots & \dots & \dots
  \enddata
\end{deluxetable*}

The final results are quoted as the fraction of simulations with a
\emph{higher} $\chi^2$ than the real data. By splitting the
simulations into two disjoint sets, and repeating the $\chi^2$
analysis for each set, we estimate that the Monte Carlo uncertainty in
the resulting significances is less than 1\%. For this reason, we quote
cases inconsistent with simulations at more than 99\% by instead
providing the actual number of simulations with higher $\chi^2$. This
allows us to distinguish between a case with $\sim$1\% significance
and one with $\sim0.1$\% significance, but still recognize the
importance of the Monte Carlo uncertainties. Further, we
conservatively never claim to obtain results with higher significance
than 99\% in any case, despite the fact that many results very likely
are far more anomalous, as would be clear by using more simulations.
 
\section{Data and simulations}

In the following, we analyze the five-year {\it WMAP} sky maps, including both
temperature and polarization. These data are available from
LAMBDA\footnote{http://lambda.gsfc.nasa.gov}, including all ancillary
data, such as beam transfer functions, noise covariance matrices and
foreground templates. The total data set spans five frequencies (23,
33, 41, 61, and 94 GHz), corresponding to the $K$, $Ka$, $Q$, $V$, and $W$ bands.

Our main objects of interest are the frequency co-added and
foreground-reduced sky maps, which are provided in the form of
HEALPix\footnote{http://healpix.jpl.nasa.gov} pixelized sky maps. The
temperature sky maps are given at a HEALPix resolution of
$N_{\textrm{side}}=512$, while the polarization maps are given at
$N_{\textrm{side}}=16$. To bring the data into a common format, we
therefore degrade the temperature component to the same resolution as
the polarization maps, simply by averaging over sub-pixels in a
low-resolution pixel. 

For temperature we consider the $Q$, $V$, and $W$ bands, and for
polarization also the $Ka$ band. This difference mirrors the official
cosmological {\it WMAP} analysis, which also uses the $Ka$ band for
polarization, but considers the same band to be too foreground
contaminated in temperature. 

Our simulations are generated as follows: we first draw a random
Gaussian CMB realization from the best-fit five-year {\it WMAP} $\Lambda$CDM
power spectrum, including multipole moments up to $\ell_{\textrm{max}}
= 1024$. This realization is then convolved with the instrumental beam
of each {\it WMAP} differencing assembly and the $N_{\textrm{pix}}=512$
HEALPix pixel window, and projected onto a HEALPix grid. For
temperature, we then add a Gaussian noise realization with
pixel-dependent variance, $\sigma^2_p = \sigma^2_0 /
N_{\textrm{obs}}(p)$, where $\sigma^2_0$ is the noise variance
per observation and $N_{\textrm{obs}}$ is the number of observations
in that given pixel. The temperature component is then degraded to
$N_{\textrm{side}}=16$.

For the polarization component, we first degrade the high-resolution
map to $N_{\textrm{side}}=16$, and then add a noise term. This is
because the polarized noise description for {\it WMAP} is given by a full
noise covariance matrix, $\mathbf{C}_{p,p'}$ at
$N_{\textrm{side}}=16$, and not as a simple $N_{\textrm{obs}}$ count
at high resolution. The correlated noise realization is generated by
drawing a vector of standard normal variates, $\eta \sim N(0,1)$, and
multiplying this vector by the Cholesky factor of the covariance
matrix, $\mathbf{n} = \mathbf{L}\eta$, where $\mathbf{C} =
\mathbf{L}\mathbf{L}^t$. 

For temperature, we adopt two different masks, namely the five-year {\it WMAP}
KQ85 and KQ75 masks. These exclude 18\% and 28\% of the sky,
respectively. For the polarization component, there is only one
relevant mask in the five-year {\it WMAP} data release, which excludes 27\% of
the sky. 

Two different polarized foreground templates are used in the
following. The first is simply the difference between the $K$ and $Ka$
bands, smoothed to an effective resolution of $10^{\circ}$ to reduce
noise, which traces synchrotron radiation. The second template is the
same starlight template as used for foreground correction in the
five-year {\it WMAP} data release, which traces thermal dust. 

\section{Results}
\label{sec:measurements}

In this section we present the two- and three-point correlation
functions computed from the polarized five-year {\it WMAP} data. Due to the
large number of available and unique correlation functions one can
form within this data set, we plot only a few selected cases, and
instead show the full set of results only in the form of tabulated
significances.

\subsection{Two-point correlations}

In Figure \ref{fig:twopt_auto}, we show a selection of two-point
correlation functions derived from the {\it WMAP} data. Specifically, all
available $Ka$, $Q$ and $V$ auto-correlations are shown, in addition to the
cross-correlations between the $Ka$ and $W$ bands. Table \ref{tab:twopt}
shows the $\chi^2$ significances for all possible two-point functions
from the available data. (Note that only unique combinations are
actually shown; empty table cells indicate that the same combinations
may be found elsewhere in the table. For example, IQ and QI are
identical for auto-correlations, but different for cross-correlations.)

Starting with the pure intensity correlations shown in the top row of
Figure \ref{fig:twopt_auto}, we recognize the by now well-known
behavior of the CMB temperature two-point function, observed both by
{\it COBE} \citep[e.g.,][]{bennett:1994} and {\it WMAP} \citep[e.g.,][]{spergel:2003}:
at small angular separations, it is slightly low compared to the
$\Lambda$CDM model, and at separations larger than $60^{\circ}$, 
very close to zero. This peculiar behavior has attracted the interest
of both experimentalists and theorists, and some have suggested that
this could be the signature of a closed topological space. However,
from the view of a simple $\chi^2$ test, this function remains
consistent with the simplest $\Lambda$CDM model at the 5\%--10\% level.

\begin{figure*}[t]
\centering
\epsfig{figure=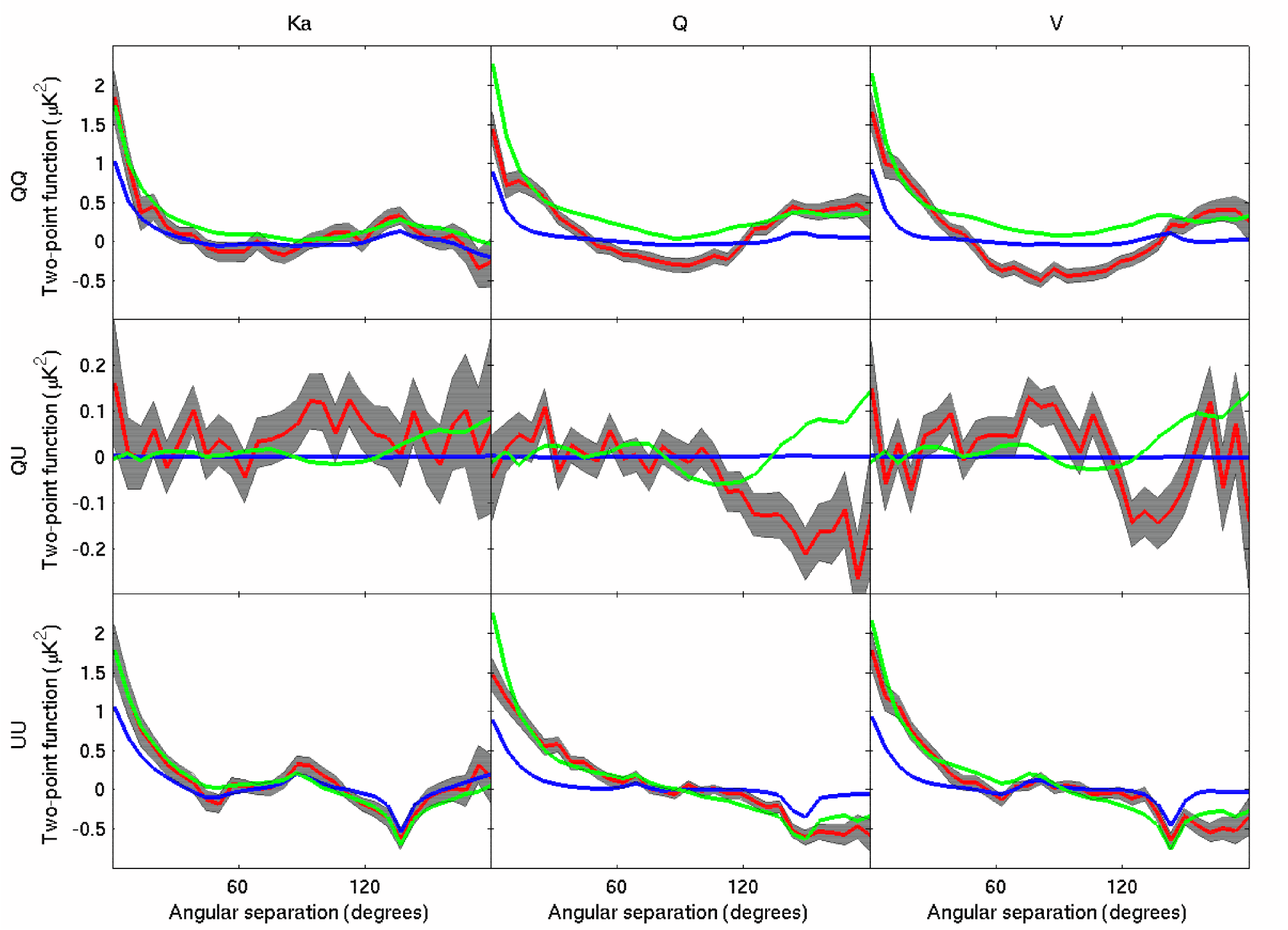,  width=\linewidth}%, totalheight=\textheight}
\caption{Template two-point functions fitted to the {\it WMAP} data two-point
functions. The red line shows the {\it WMAP} data, with the gray area being the
1 $\sigma$ confidence intervals. The blue line shows the the mean correlation 
function of the simulations, while the green line shows the mean value plus
the foreground contribution.}
\label{fig:amp_tp}
\end{figure*}

Next, looking at the temperature-polarization (I-Q/U)
cross-correlations we see that the overall behavior of the $Q$-, $V$, and
$Ka \times W$ band functions are quite different. This indicates that the
{\it WMAP} data still are too noise dominated to extract high-sensitivity
cosmological information from individual bands. This is also reflected
in the second and third columns of Table \ref{tab:twopt}, where there
is a significant scatter between the different results. However, we do
see that all results are in good agreement with the expectations,
indicating that the noise model is satisfactory.

The three bottom rows of Figure \ref{fig:twopt_auto} show the pure
polarization correlation functions, which are the main target in this
paper. And here we see several interesting features. First, we
recognize a sharp feature in the UU correlation function at
$\sim141^{\circ}$. This is the angular separation between the A- and
B sides of the differential {\it WMAP} detectors, and it was also seen in
the pure noise temperature two-point correlation function in the
first-year {\it WMAP} data \citep{eriksen:2005}. 

However, even more interesting is the overall very pronounced
large-scale excesses seen in the QQ and UU functions: both the $Q$- and
$V$-band functions lie mostly within the $2\sigma$-$3\sigma$ confidence regions,
and the overall agreement with the simulations appears quite poor. 

Again, this is strongly reflected in Table \ref{tab:twopt}: all QQ
correlations are anomalous at more than 99\% confidence, and all UU
correlations at more than 98\% confidence. In the case of the
$W-$band, which happens to be the most anomalous of any band, we have
good reasons to expect such behavior. This band has a significantly
higher $1/f$ knee frequency than any other band \citep{jarosik:2003},
with the $W$4 differencing assembly having the highest. As a result,
the {\it WMAP} team has chosen not to use this band for cosmological
analysis in polarization. But the anomalous behavior of the $Q$ and
$V$ bands is a priori not expected; these are used for cosmological
analysis by the {\it WMAP} team and should be clean.

We have also generated an ensemble of noise-only simulations, and
computed the two-point functions from these. The corresponding
$\chi^2$ fractions are listed in parentheses in Table \ref{tab:twopt}
for the pure polarization modes. Here we see that, generally speaking,
the noise-only hypothesis performs almost, but not quite, as well as
the signal-plus-noise hypothesis. This is simply due to the fact that
the {\it WMAP} polarization data are strongly dominated by the correlated
noise, and it is very difficult from a correlation function
point of view to distinguish between a small signal component and a
large-scale noise fluctuation.

\subsubsection{Foreground cross-correlations}

The anomalous behavior seen in the $Q$- and $V$-band two-point functions
clearly needs an explanation. And typically, when such unexpected
behavior is observed, one of the the first issues to consider is
residual foregrounds.

To check whether this may be a relevant issue, we first simply compute
the cross-correlation functions between each of the three frequency
bands ($Ka$, $Q$, and $V$) and the two available foreground templates
(synchrotron and dust). This is done both for the real {\it WMAP} data and
the simulated ensembles, and the agreement between the (pure CMB $+$
noise) simulations and the {\it WMAP} data is again quoted in terms of a
$\chi^2$ fraction.

The results from these calculations are as follows: for the $Ka$ band,
we find that the $\chi^2$ significances are 0.51 and 0.24 for
synchrotron and dust, respectively, indicating no significant
foreground detection in this case. However, for the $Q$ band the
corresponding numbers are 0.05 and $\sim0.005$, corresponding to
correlations statistically significant at $2\sigma$ and $\sim3\sigma$,
respectively. For the $V$ band, the numbers are 0.05 and 0.02,
significant at $2\sigma$ or more.

To study these foreground correlations further, we compute the
two-point functions from the foreground templates directly, and fit
these to the observed correlation functions with a single free
amplitude for each template, $A_{\textrm{s}}$ and $A_{\textrm{d}}$, by
minimizing
\begin{eqnarray*}
  \chi^2_{\textrm{fg}} &=& \sum_{bb'} \left(C_{\textrm{obs}}(b) -
  C_{\textrm{sim}}(b) - A_{\textrm{s}} C_{\textrm{s}}(b) -
  A_{\textrm{d}} C_{\textrm{d}}(b)\right)\Sigma^{-1}_{\textrm{sim}}(b, b')\\
  &\times& \left(C_{\textrm{obs}}(b') -
  C_{\textrm{sim}}(b') - A_{\textrm{s}} C_{\textrm{s}}(b') -
    A_{\textrm{d}} C_{\textrm{d}}(b')\right).
\end{eqnarray*}
Here, $C_{\textrm{sim}}(b)$ is the
correlation function mean and $\Sigma_{\textrm{sim}}(b, b') = \bigl<(C(b) -\mu(b))(C(b')-\mu(b'))\bigr>$ is the covariance matrix,
both quantities obtained from the noise simulations.
The indices $b, b'$ run over all possible pure polarization
two-point bins (i.e., both angular bins and QQ, QU, and UU
correlations). 

The resulting best-fit amplitudes from this calculation for $Ka$-, $Q$-
and $V$ bands are shown in Table \ref{tab:amp_marg_conf}. Here, we again
see that the $Ka$-band amplitudes are generally significantly lower than
those of the $Q$ and $V$ bands. (Note that the uncertainties quoted in
this table only include statistical errors, not systematic errors. The
significances should therefore not be considered as true detection
levels, but are only suggestive.)

\begin{deluxetable}{ccccc}
%  \tablewidth{\linewidth}
  \tablecaption{Amplitude confidence intervals \label{tab:amp_marg_conf}}
  \tablecomments{
  Mean values and confidence intervals for the best-fit amplitudes of the
  $K-Ka$ and dust template two-point functions relative to the {\it WMAP} data
  two-point functions.}
  \tablecolumns{3}
  \tablehead{& \multicolumn{2}{c}{Synchrotron} &
    \multicolumn{2}{c}{Dust}\\ Band & $A_{\textrm{s}}$ &
    $\sigma_{sig}$ & $A_{\textrm{d}}$ & $\sigma_{sig}$}
  \startdata
  $Ka$ & 0.0038$\pm 0.0013 $ & 2.9 & 0.0115$\pm 0.0057$  & 2.0 \\
  $Q$ & 0.0052$\pm 0.0010$   & 5.2 & 0.0256$\pm 0.0042$  & 6.1 \\ 
  $V$ & 0.0063$\pm 0.0011$   & 5.7 & 0.0197$\pm 0.0047$  & 4.2
  \enddata
\end{deluxetable}

In Figure \ref{fig:amp_tp}, the fitted correlation functions are
compared to the observed functions. First, the red curve shows the
functions derived from the actual {\it WMAP} data, with gray bands
indicating the $1\sigma$ uncertainties derived from simulations. The
blue curve shows the mean correlation function of the simulations,
$C_{\textrm{sim}}(b)$, and finally, the green curve shows the same,
but with the foreground contribution added in, $C_{\textrm{fg}}(b) =
C_{\textrm{sim}}(b) + A_{\textrm{s}} C_{\textrm{s}}(b) +
A_{\textrm{d}} C_{\textrm{d}}(b)$. Clearly, the latter matches the
real data far better than the pure CMB-plus-noise hypothesis, with the
best match seen in the UU correlation function.

\subsection{Three-point correlations}

\begin{deluxetable*}{ccccccccc}[t]
  \tablewidth{\linewidth}
  \tablecaption{Three-point function results
  \label{tab:threept}}
  \tablecomments{Statistical significances of the polarization-only three-point
  functions computed from the {\it WMAP} five-year co-added $Ka$, $Q$ and $V$ maps. The Monte Carlo uncertainty is 1\%.
  See Table \ref{tab:twopt} for feature explanation.
  }
  \tablecolumns{9}
  \tablehead{& \multicolumn{8}{c}{Correlation} \\ Frequency & QQQ & QQU & QUQ & QUU & UQQ & UQU & UUQ & UUU}
  \startdata
  \cutinhead{Equilateral functions}
   $Ka$x$Q$x$V$ & 0.04 (0.03) & 0.31 (0.26) & 18$^*$ (4$^*$) & 0.04 (0.03) & 0.16 (0.13) & 0.01 (416$^*$) & 0.07 (0.05) & 0.21 (0.17) \\
   $Ka$x$V$x$Q$ & 0.45 (0.40) & 0.10 (0.08) & 0.07 (0.05) & 0.01 (446$^*$) & 359$^*$ (241$^*$) & 0.35 (0.31) & 0.01 (0.01) & 0.11 (0.09) \\
  \cutinhead{Collapsed functions}
   $Ka$x$Q$x$V$ & 0.49 (0.45) & 0.01 (0.01) & 0.23 (0.18) & 0.03 (0.02) & 0.63 (0.56) & 0.01 (384$^*$) & 0.21 (0.18) & 0.43 (0.38) \\
   $Ka$x$V$x$Q$ & 323$^*$ (240$^*$) & 0.01 (0.01) & 2$^*$ (1$^*$) & 0.05 (0.03) & 0$^*$ (1$^*$) & 0.11 (0.09) & 0.05 (0.03) & 17$^*$ (11$^*$) \\
   $Q$x$Ka$x$V$ & 0.01 (0.01) & 0.28 (0.25) & 0.01 (329$^*$) & 0.17 (0.14) & 0.01 (298$^*$) & 231$^*$ (179$^*$) & 0.07 (0.05) & 0.15 (0.12) \\
   $Q$x$V$x$Ka$ & 0.05 (0.04) & 0.09 (0.07) & 0.01 (0.01) & 0.10 (0.08) & 0.04 (0.03) & 0.07 (0.06) & 0.12 (0.09) & 0.12 (0.09) \\
   $V$x$Ka$x$Q$ & 0.21 (0.17) & 0.04 (0.03) & 180$^*$ (5$^*$) & 0$^*$ (2$^*$) & 3$^*$ (2$^*$) & 127$^*$ (79$^*$) & 80 (45$^*$) & 0.01 (456$^*$) \\
   $V$x$Q$x$Ka$ & 0.01 (0.01) & 0.03 (0.02) & 0.58 (0.51) & 439$^*$ (301$^*$) & 0.01 (0.01) & 0.38 (0.33) & 0.18 (0.13) & 0.05 (0.04) 
  \enddata
\end{deluxetable*}

We now turn to the three-point correlation functions derived from the
five-year polarized {\it WMAP} data. However, we note that the
three-point function is primarily used in the literature as a test of
non-Gaussianity. In the present case, this will not be case, since we
have already seen that the data appear to be foreground contaminated,
and even the two-point correlation function fails a standard $\chi^2$
test. The following three-point analysis will therefore essentially
only be a consistency check of the above results, and a demonstration
of the general procedure, to be further developed in the future, when
cleaner data become available.

\begin{figure*}[t]
\centering
\epsfig{figure=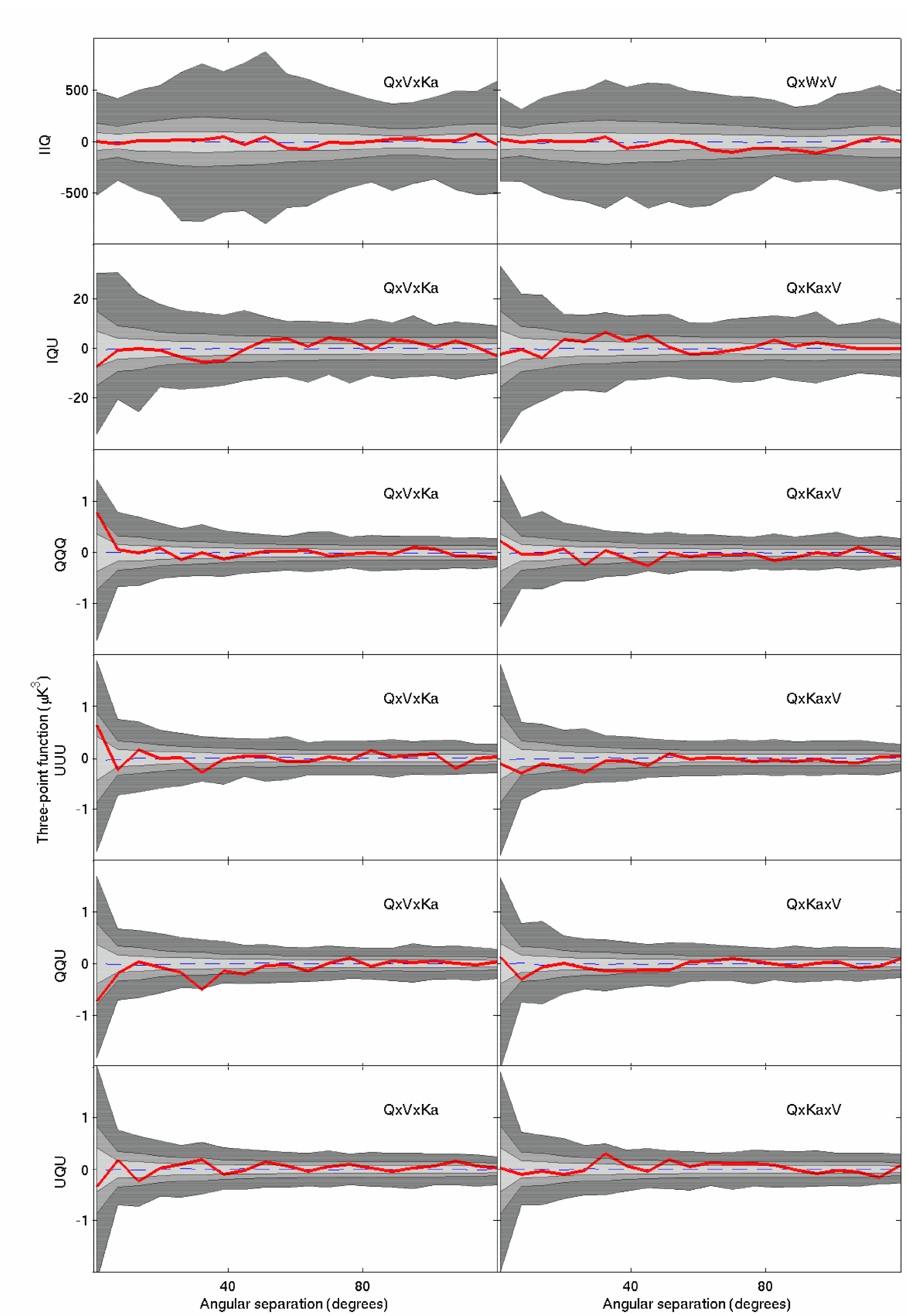, width=0.9\linewidth}%, totalheight=0.41\textheight}
\caption{Equilateral three-point functions, computed with the KQ85 mask, as functions of angular
separation on the sky. Only a representative selection is shown. 
The red line is the three-point function
computed from the {\it WMAP} five-year co-added maps. 
The light, medium and dark gray shaded areas correspond to 1$\sigma$, 
2$\sigma$, and 3$\sigma$, respectively. The dashed blue line is the maximal
point of the correlation histograms. 
Each row corresponds to a specific
combination of modes, while the frequency bands involved are marked on each
plot.}
\label{fig:threept_eq}
\end{figure*}

We consider only cross-correlations of pure polarization modes for the
cosmologically interesting frequency channels. This leaves us with
only eight independent correlation function modes (QQQ, QUQ, UQQ, UUU
etc.), and only one frequency combination ($Ka x Q x V$). Further,
we consider only two general types of three-point configurations,
namely, equilateral and pseudo-collapsed triangles. The latter is
defined such that two legs of the triangle are of equal length, while
the third leg spans an angular distance of less than $12^{\circ}$. 

%\begin{deluxetable}{ccc}
%  \tablewidth{\linewidth}
%  \tablecaption{Reduced statistical significances \label{tab:sigsum}}
%  \tablecomments{The significance of the statistical significances
%  for the CMB hypothesis and the noise-only hypothesis. See table
%  \ref{tab:twopt} for feature explanation.}
%  \tablecolumns{3}
%  \tablehead{& CMB & Noise-only} 
%  \startdata
%  All &\ldots & \ldots \\
%  II and I(Q/U) correlations & \ldots & \ldots \\
%  Pure QU correlations & \ldots & \ldots
%  \enddata
%\end{deluxetable}

The results from these calculations are tabulated in Table
\ref{tab:threept}, and a few arbitrarily chosen modes of the
equilateral three-point functions are plotted in Figure
\ref{fig:threept_eq}. (Plotting all functions would take too much
space, without adding any particular new insights.)

Once again, we see that the $\chi^2$'s obtained from the real data are
high, but generally not quite as striking as in the two-point
case. This is at least partly due to the fact that the $Ka$ band is
involved in all configurations, as well as the fact that the
three-point function is always ``noisier'' than the two-point
function. We also see that the noise-only hypothesis always has higher
$\chi^2$'s than the signal-plus-noise hypothesis, again following the
two-point behavior. 

It is difficult to interpret these results much further, given that
there are clear problems already at the two-point level. We therefore
leave a full interpretation of the three-point function to a future
publication, when cleaner data, either from {\it WMAP} or {\it
Planck}, have been made available.

\section{Conclusions}
\label{sec:conclusions}

We compute for the first time the polarized two- and three-point
correlation functions from the five-year {\it WMAP} data, with the initial
motivation of constraining possible non-Gaussian signals on large
scales in the CMB polarization sector. However, our main result is a
detection of a signature consistent with residual foregrounds in both
the $Q$- and $V$-band data, significant at $2\sigma$-$3\sigma$. 

This should not come as a complete surprise, though, as a similar
result was found in the three-year {\it WMAP} data through a direct template fit
approach by \citet{eriksen:2007}. They used a Gibbs sampling
implementation to estimate the joint CMB power spectrum and foreground
template posterior, and found non-zero template amplitudes at similar
significance levels when marginalizing over the CMB spectrum. Given the
results found in the present paper, this still appears to be an
important issue for the five-year {\it WMAP} data release. Resolving this
question is of major importance for the CMB community, since many
cosmological parameters depend critically on the large-scale {\it WMAP}
polarization data, most notably the optical depth of reionization,
$\tau$, and the spectral index of scalar perturbations,
$n_{\textrm{s}}$. 

The question of primordial non-Gaussianity in CMB polarization data
can not be meaningfully addressed as long as this issue is
unresolved. The three-point correlation function results shown in this
paper therefore mostly serve as a demonstration of the approach, and
as a cross-check on the two-point results. The three-point function
will obviously become a more important and independent quantity in the
future, when higher-fidelity polarization data become available.

\begin{acknowledgements}
We acknowledge use of the
HEALPix\footnote{http://www.eso.org/science/healpix/} software
\citep{healpix} and analysis package for deriving the results in
this paper, and use of the Legacy Archive for Microwave Background
Data Analysis (LAMBDA). H.K.E. and P.B.L. acknowledge financial support from
the Research Council of Norway. The computations presented in this
paper were carried out on Titan, a cluster owned and maintained by the
University of Oslo and NOTUR.
\end{acknowledgements}

\appendix
\section{Precomputing configuration tables}
In order to compute the two-and-three point functions, we have utilized the
method
introduced by \citet{eriksen:2004b}. 

The basic idea of this method is that given the resolution and the
mask of the map, and the desired number of distance bins, the
configurations that correspond to an angle $\theta$ is uniquely
determined.  One may then do the work of finding these configurations
only once, saving them in large tables. Each table then corresponds to
a certain bin $k$, and the first row in each table contains all the
unmasked pixels of the map. Each column contains all pixels at a
distance given by $kd\theta$ from the pixel in the first element of
the column. By moving through a column in such a table, one then
traces out a circle of radius $kd\theta$.  Thus, by saving the
configurations in tables, one pays the initialization costs only once.

An additional, powerful feature of this approach becomes apparent when
computing correlation functions of higher order than 2. If one thinks of these
tables as compasses, tracing out circles of a given length for each pixel, it
should be clear that, just as one can construct, e.g., triangles using compasses,
one can construct triangles on the sphere using these tables. If the edges of 
the desired triangle is of lengths $\theta$, $\alpha$ and $\beta$, one begins
with the $\theta$ table, selects a pixel from the upper row, and selects a
second pixel from the column below this pixel. One then has the baseline of
the triangle. To construct a triangle with a compass when the baseline is
given, one would adjust the compass so that it spans a length equal to the 
length of the second edge of the triangle, and draw a circle with this radius
with one end of the baseline as the center of the circle. Then, one would
again adjust the compass so that it spans the desired third length of the
triangle, and draw a circle around the other end of the baseline. Then, one
would check whether the two circles drawn intercept at any point(s). These
points would then be the third vertex of the triangle. Carrying this analogy
to our tables, our two already chosen pixels are the ends of the baseline 
around which to draw our circles. We 'adjust our compass' by finding pixel 1 in
the first row of table $\alpha$, we 'draw a circle around it' by looping over
all pixels in the column
below it, and do the same for pixel 2, expect that we now look in table
$\beta$ . By checking whether there are any common pixels in these 'circles', 
we find the triangle(s) whose edges have the desired lengths. 

Since any $N$-point polygon can be reduced to triangles, higher-order
correlation functions can also be computed using this method.

\end{document}